\documentclass[conference]{IEEEtran}
\IEEEoverridecommandlockouts
\usepackage{cite}
\usepackage{amsmath,amssymb,amsfonts}
\usepackage{algorithmic}
\usepackage{graphicx}
\usepackage{textcomp}
\usepackage{xcolor}
\usepackage[utf8]{inputenc} % use UTF8 encoding
\usepackage{kotex} % use KoTeX package for Korean
\usepackage{physics} % 브라켓 및 수식을 위해 사용
\usepackage{amsmath} % 행렬
\usepackage{qcircuit} % Quantum circuit을 그리기 위해 사용

\def\BibTeX{{\rm B\kern-.05em{\sc i\kern-.025em b}\kern-.08em
    T\kern-.1667em\lower.7ex\hbox{E}\kern-.125emX}}
\begin{document}

\title{A Tutorial on Quantum Convolutional Neural Networks (QCNN)}

\author{\IEEEauthorblockN{$^{\circ}$Seunghyeok Oh, $^{\dag}$Jaeho Choi, and $^{\circ}$Joongheon Kim}
\IEEEauthorblockA{$^{\circ}$School of Electrical Engineering, Korea University, Seoul, Republic of Korea \\
$^{\dag}$School of Computer Science and Engineering, Chung-Ang University, Seoul, Republic of Korea\\
E-mails: 
\texttt{seunghyeokoh@korea.ac.kr}, \texttt{jaehochoi2019@gmail.com}, 
\texttt{joongheon@korea.ac.kr}}
}

\maketitle

\begin{abstract}
Convolutional Neural Network (CNN) is a popular model in computer vision and has the advantage of making good use of the correlation information of data. However, CNN is challenging to learn efficiently if the given dimension of data or model becomes too large. Quantum Convolutional Neural Network (QCNN) provides a new solution to a problem to solve with CNN using a quantum computing environment, or a direction to improve the performance of an existing learning model. The first study to be introduced proposes a model to effectively solve the classification problem in quantum physics and chemistry by applying the structure of CNN to the quantum computing environment. The research also proposes the model that can be calculated with O(log(n)) depth using Multi-scale Entanglement Renormalization Ansatz (MERA). The second study introduces a method to improve the model's performance by adding a layer using quantum computing to the CNN learning model used in the existing computer vision. This model can also be used in small quantum computers, and a hybrid learning model can be designed by adding a quantum convolution layer to the CNN model or replacing it with a convolution layer. This paper also verifies whether the QCNN model is capable of efficient learning compared to CNN through training using the MNIST dataset through the TensorFlow Quantum platform.
\end{abstract}

\section{Introduction}

Quantum computers are emerging as a new solution to problems not solved by classical computers.
% 양자 컴퓨터는 기존의 컴퓨팅 방법으로 해결하지 못한 문제들의 새로운 해결책으로 각광받고 있다. 
Quantum computers provide a computing environment that is different from classical computers.
% 양자 컴퓨터은 기존 컴퓨터와는 완전히 다른 방식의 컴퓨팅 환경을 제공한다.
In particular, quantum computers can use superposition and entanglement, not seen in classical computing environments, and obtain powerful performance using parallelism between qubits\cite{bravyi2018quantum}.
% 특히, 양자 컴퓨터에서는 superposition, entanglement등의 고전 컴퓨팅 환경에서는 볼 수 없었던 방식을 이용할 수 있으며, qubit들간의 parallism을 이용하여 강력한 퍼포먼스를 얻을 수 있다.
Through these advantages, the quantum computer is considered new solutions to algorithmic problems that cannot be easily solved. 
% 이 특징들을 이용하여, 기존에 쉽게 해결할 수 없는 알고리즘 문제들의 새로운 해결책으로 주목받고 있다.
Also, in the field of machine learning, various studies applying quantum computing models are in progress.
% 또한, 머신러닝 분야에서도 양자컴퓨팅 모델을 적용한 다양한 연구들이 진행중이다.
There are Variational Quantum Eigensolver (VQE) \cite{peruzzo2014variational, McClean_2016} and Quantum Approximate Optimization Algorithm (QAOA)\cite{farhi2014quantum, hadfield2019quantum}, which provide new ways to solve the problems of physics and chemistry in complex structures or NP-hard algorithm problems.
% 예를 들어, Variational Quantum Eigensolver(VQE)와  Quantum Approximate Optimization Algorithm(QAOA) 등이 있으며, 이들은 복잡한 구조의 물리,화학 분야의 문제나, NP-hard 알고리즘 문제를 해결하기 위한 새로운 방법을 제공해주었다.
Besides, as optimization using gradient descent method in quantum devices has been studied, learning quantum machine learning using hyperparameters can be performed efficiently\cite{schuld2019evaluating, stokes2020quantum}.
% 또한, 양자 디바이스에서도 gradient descent 방법을 이용한 효율적인 학습방법이 연구되면서, 하이퍼 파라매터를 이용한 학습을 할 수 있게 되었다.

Convolutional Neural Network (CNN), among many classification models, has shown very high performance in computer vision\cite{krizhevsky2012imagenet}.
% Convolution Neural Network(CNN)은 많은 분류 모델중, 컴퓨터 비전 분야에서 매우 높은 성능을 보여주었다
Images that reflect the real world, such as photographs, have a very high correlation between surrounding pixels.
% 사진 등의 현실 세계를 반영하는 이미지들은 일반적으로 공간 정보나, 주변 픽셀간에 매우 높은 correlation을 갖게 된다.
The fully-connected layer, basic model in deep learning, showed strong performance in machine learning, but there is no way to keep the correlation.
% 딥러닝에서의 기본 모델인 fully-connected layer 은 머신 러닝에서 강력한 성능을 보여 주지만 위의 정보를 유지할 방법은 없다.
On the other hand, CNN can maintain correlation information directly, resulting in better performance evaluation.
% 반면 CNN은 이러한 주변의 correlation 정보를 잘 보존하기 때문에 더 좋은 성능평가를 얻을 수 있다.

CNN mainly proceeds by stacking the convolution layer and the pooling layer. 
% CNN은 주로 Convolution layer와 Pooling layer를 겹겹히 쌓는 방식으로 진행한다.
The convolution layer finds new hidden data by linear combinations between surrounding pixels.
% Convolution layer는 주변 픽셀간의 linear combination으로 새로운 hidden data를 찾아낸다.
The pooling layer reduces the size of the feature map, reducing the resources required for learning and avoiding overfitting.
% Pooling layer는 feature map의 크기를 축소시켜, 학습에 요구되는 리소스를 줄일 수 있고 overfitting을 막을 수 있다.
When the data size is sufficiently reduced by repeatedly applying these layers, the classification result is obtained using the fully connected layer.
% 두 레이어들을 반복적으로 적용해서 충분히 데이터 크기를 줄이게 되면, fully connected layer를 적용하고, 분류 결과를 예측한다.
The loss between the acquired label and the actual label can train the model using a gradient descent method or other optimizers for better results.
% 이렇게 얻어진 label과 실제 label간의 MSE를 구하고, 이를 gradient descent 방식, 또는 이를 응용한 optimizer를 이용하여 학습을 할 수 있다.

However, many problems that exist in the real world are still hard to solve with classic machine learning methods.
% 하지만, 현실 세계에 존재하는 많은 문제들이 아직 고전적인 머신러닝 방법으로는 해결하기 어렵다.
The quantum physics problem defined in the many-body Hilbert space requires converting these data into classical computer data to apply machine learning techniques.
% many-body Hilbert space에서 정의된 양자 물리 문제는 머신러닝 적용시키기 위해서는 이 데이터들을 고전적 컴퓨터 데이터로 변환해야한다.
As the size of the system increases, the size of the data increases exponentially, making it difficult to solve effectively even with a machine learning method.
% 하지만, 시스템의 크기가 커질수록, 데이터의 크기가 폭발적으로 커지기 때문에, 고전적 환경에서의 머신러닝 방법으로도 효율적으로 풀기 힘들어진다.
In addition to the above case, other alternatives when data and models are no longer efficiently processed in existing computing environments.
% 그밖에도, 데이터와 모델이 커지면서 더이상 기존 컴퓨팅 환경에서 효율적으로 처리하지 못하는 경우에는 다른 대안이 필요하다.

Many studies have appeared to solve these problems with the Quantum Convolutional Neural Network (QCNN) using the quantum computing system and the CNN model together.
% 최근 양자 컴퓨팅 시스템과 CNN모델을 함께 이용한 Quantum Convolution Neural Network(QCNN)으로 이러한 문제들을 해결하려는 많은 연구들이 등장하였다.
There is an approach to apply the CNN structure itself to a quantum system to efficiently solve quantum physics problems and an approach to improve performance by adding a quantum system to problems previously solved by CNN.
% 대표적으로, 현실의 양자 물리를 효율적으로 해결하기 위해 CNN 구조자체를 양자 시스템에 적용하려는 방법과,기존에 CNN에서 해결 하던 문제들에 양자 시스템을 추가하여 성능을 향상시키는 방법이 있다.
This paper introduces these studies.
% 이 papaer에서는 이러한 연구들에 대해 소개하고자 한다.

\begin{figure}[t]
    \centering
    \includegraphics[width=\linewidth]{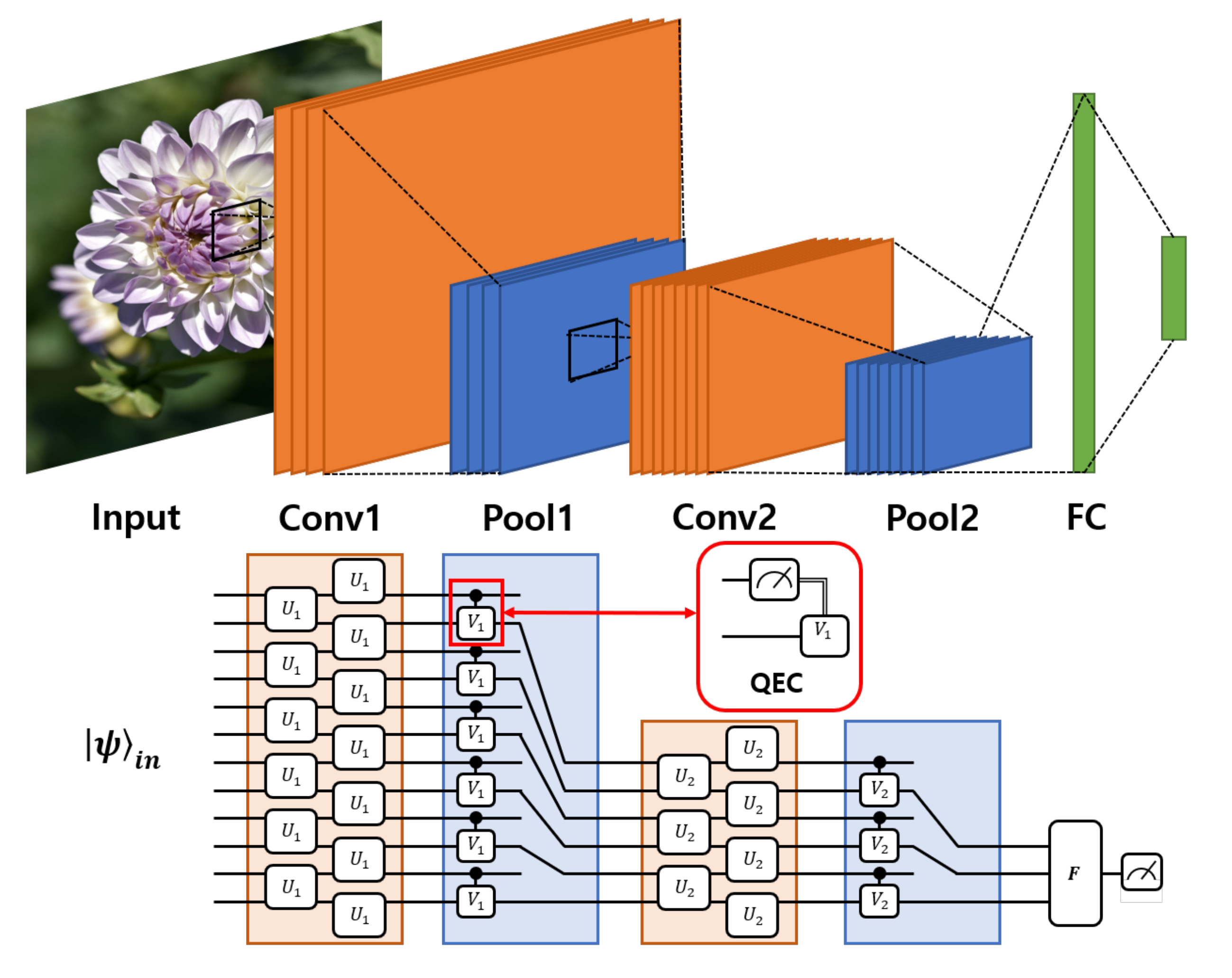}
    \caption{Simple example of CNN and QCNN. QCNN, like CNN, consists of a convolution layer that finds a new state and a pooling layer that reduces the size of the system. QEC can be additionally applied by using measurement instead of controlled-gate in the pooling layer\cite{cong2019quantum}.}
    % CNN과 QCNN의 간단한 예시.
    % QCNN은 CNN과 마찬가지로 새로운 state를 찾아내는 convolution layer와, system의 크기를 줄이는 pooling layer로 구성되어있다.
    % 또한, Pooling layer에서 controlled-gate 대신에 measurement를 이용해 QEC를 추가로 적용할 수 있다.
    \label{fig1}
\end{figure}

\section{QCNN Using The Structure of CNN}

QCNN proposed by \textit{Cong} extends the main features and structures of the existing CNN to quantum systems\cite{cong2019quantum}. %%% 이름
% Cong이 제시한 QCNN는 기존 CNN의 주요 특징과 구조를 양자 시스템에 확장하였다.
% 양자물리, 화학 문제들은 입자간의 상호작용이 크게 고려되기 때문에, CNN 방식의 모델은 매우 효과적이다. 
In moving the quantum physics problem defined in the many-body Hilbert space to the classical computing environment, the data size exponentially increases according to the system size, so it is not suitable to solve efficiently.
% 고전적 컴퓨팅 환경으로는 many-body Hilbert space에서 정의된 실제 자연의 양자물리 문제를 풀기에는 데이터 사이즈가 시스템 크기에 따라 exponential하게 커지기 때문에 효율적으로 풀기 부적합하다.
In a quantum environment, the data can express using qubits, so the problem can be avoided by applying a CNN structure to a quantum computer.
% 양자환경에서는 이 데이터들을 큐빗에 그대로 표현할 수 있기 때문에, 양자 컴퓨터에 CNN 구조를 적용해서 이 문제를 해결할 수 있다.
This section briefly introduces the structural design of this QCNN.
% 이번 섹션에서는 이 QCNN의 구조 설계에 대해서 간단히 소개한다.

The model of QCNN applies the convolution layer and the pooling layer, which are the main features of CNN, to quantum systems, as shown in Fig.~\ref{fig1}.
% QCNN의 모델은 CNN의 주요 특징인 convolution layer와 pooling layer를 양자 시스템에 적용한다.
The concept proceeds as follows:
% 주요 컨셉은 다음과 같다.

\begin{enumerate}
    \item The convolution circuit finds the hidden state by applying multiple qubit gates between adjacent qubits.
    % a) Convolution circuit은 인접 큐빗간의 multiple qubit gate를 적용하여 hidden state를 찾아낸다.
    \item The pooling circuit reduces the size of the quantum system by observing the fraction of qubits or applying 2-qubit gates such as CNOT gates.
    % Pooling circuit은 특정 비율의 qubit을 관측하거나, CNOT gate 등의 2-qubit gate를 적용하여 모델에서 제외시켜 양자 시스템의 크기를 축소시킨다.
    \item Repeat the convolution circuit and pooling circuit defined in $1)$-$2)$.
    % $1)$-$2)$에 정의된 convolution circuit과 pooling circuit 를 반복적으로 적용한다.
    \item When the size of the system is sufficiently small, the fully connected circuit predicts the classification result.
    % system의 크기가 충분히 작아지면, fully connected circuit을 통해분류 결과를 예측한다.
\end{enumerate}
% 삭제)이 모델은 양자 시스템의 크기는 exponential하게 축소시키고, 이는 parameter의 수를 줄여 더 빠른 최적화를 가능하게 해준다.

The model used to satisfy this structure is typically Multi-scale Entanglement Renormalization Ansatz (MERA)\cite{vidal2008class}.
% 이러한 구조를 만족하시키기 위해 사용되는 모델은 대표적으로 Multi-scale Entanglement Renormalization Ansatz (MERA)가 있다.
MERA is a model designed to simulate many-body state quantum systems efficiently.
% MERA는 기존에 many-body state quantum system을 효율적으로 시뮬레이션 하기위해 고안된 모델이다.
% TODO: MERA에 대해 한줄 더 추가
At this time, MERA exponentially increases the size of the quantum system for each depth by adding qubits of $\ket{0}$.
% 이때, MERA는 각 depth마다 |0> 상태의 qubit을 추가하여 양자 시스템의 크기를 exponential하게 늘려간다.
QCNN uses this MERA in the reverse direction.
% QCNN은 이 MERA를 역방향으로 이용한다.
The reversed MERA reduces the size of the quantum system exponentially from the given data, which is suitable as a model of QCNN.
% 그 경우, 주어진 데이터로부터 양자시스템의 크기를 exponential하게 줄일 수 있기 때문에 QCNN의 모델로 적합하다.
% TODO: MERA를 이용한 구조 이미지 -> 이미지가 너무 많아서 넣지 않아도 될까..?

The QCNN model proposed by \textit{Cong} suggests an additional performance improvement through the Quantum Error Correction (QEC) to this MERA model\cite{preskill1998lecture}.
% Cong이 제시한 QCNN 모델은 이 MERA 모델에 Quantum Error correction (QEC)을 통해 추가적인 성능향상의 뱡향을 제시한다.
There is a representative state $\ket{\psi}$ for each label in MERA.
% MERA에서 각 레이블에 따라 대표하는 상태 \psi>가 존재한다.
Since QCNN uses the reverse direction of MERA, if its $\ket{\psi}$ is given as input data, the corresponding label can be obtained as a definitive solution.
% QCNN은 MERA의 역방향을 이용하기 때문에, 그 \psi>가 input data로 주어지게 되면 그에 해당하는 레이블을 결정적인 해로 얻을 수 있다.
On the other hand, if $\ket{\psi'}$ that cannot be generated in MERA is given as input data, QCNN cannot obtain a definitive solution.
% 반면, MERA에서 생성될 수 없는 \psi'>가 input data로 주어지게 될 경우에는 결정적인 해를 얻을 수 없다.
This problem can be corrected and solved by applying QEC to give additional degrees of freedom.
% 이 문제를 QEC를 적용하여 추가적인 자유도를 부여해 보정, 해결할 수 있다.

When the data given to QCNN is $\ket{\psi}$, the result measured in the pooling layer should be $\ket{0}$ the same as the newly given state in MERA. 
% 만약 QCNN에 주어진 데이터가 \psi>인 경우에는, pooling layer에서 측정되는 결과는 MERA에서 새롭게 주어지는 상태와 같은  \0>이어야 한다.
On the other hand, if $\ket{\psi'}$ which MERA cannot generate is given as input data, there is a possibility that it will be $\ket{1}$ in the measured result. 
% 반면 MERA에서 생성될 수 없는 상태인 \psi'>가 입력 데이터로 주어질 경우, 측정되는 결과중에 1일 가능성이 존재하게 된다.
Using them, if $\ket{1}$ is measured, an additional gate is applied to the surrounding qubits to correct the result. 
% 이 결과를 이용해, 만약 1이 관측 될 경우에 주변의 큐빗에 추가적인 gate를 적용해 결과를 보정한다.
The method can give better performance through additional deterministic measurement outcomes.
% 이 방법을 이용하여 추가적인 결정적인 해를 통해 성능향상을 얻는다. 

\section{QCNN for Image Classification}

Image classification is one of the most applied fields in neural networks, such as CNN.
% CNN 등의 뉴럴 네트워크에서 가장 많이 적용되는 분야 중 하나는 image classification이다.
Quantum computers have potent advantages in terms of superposition and parallel computation.
% 양자컴퓨터는 superposition과 parallel computation쪽에서 강력한 장점을 갖고 있다.
Quantum Convolutional Neural Network proposed by \textit{Henderson} applies quantum environments in CNN to improve the performance of CNN\cite{henderson2020quanvolutional}.
% Henderson의 Quanvolutional Neural Network에서는 CNN에 양자 환경을 적용해 CNN의 퍼포먼스를 향상시키고자 한다.
This section briefly introduces the research that suggested how to apply a quantum computing system to CNN.
% 이번 섹션에서는 CNN에 quantum compputing system을 적용하는 방법을 제시한 연구에 대해 간단히 소개해본다.

The quantum convolution layer defines a layer that behaves like a convolution layer in a quantum system.
% Henderson이 제시한 Quanvolution layer는 Convolution layer와 같은 동작을 하는 레이어를 양자 시스템에 정의한다.
The quantum convolution layer applies a filter to the input feature map to obtain feature maps composed of new data.
% Quanvolution layer는 convolution layer와 비슷하게 input feature map을 filter를 적용하여 새로운 data로 구성된 feature map을 얻는다.
However, the quantum convolution layer uses a quantum computing environment for filter operation, unlike the convolution layer.
% 단, Quanvolution layer는 Convolution layer와 다르게 filter 연산을 양자 컴퓨팅 환경을 이용한다.

Quantum computers have the advantages of superposition and parallel computation that do not exist in classical computing, which can reduce the learning time and evaluation time.
% 양자컴퓨터는 고전 컴퓨팅에는 존재하지 않는 superposition과 parallel computation의 강력한 장점을 갖고 있기 때문에, 학습 속도를 줄일 수 있다.
However, existing quantum computers are still limited to small quantum systems.
% 하지만, 현존하는 양자컴퓨터는 아직 소규모의 양자 시스템으로 제한되어있다.
The quantum convolution layer does not apply the entire image map to a quantum system at once, but processes it as much as the filter size at a time, so small quantum computers can construct the quantum convolution layer.
% Quanvolution은 이미지 전체를 한번에 양자 시스템에 적용하는게 아닌, 한번에 filter 크기만큼 처리하기 때문에, 소규모양자 컴퓨터에서도 사용할 수 있다.

\begin{figure}[t]
    \centering
    \includegraphics[width=\linewidth]{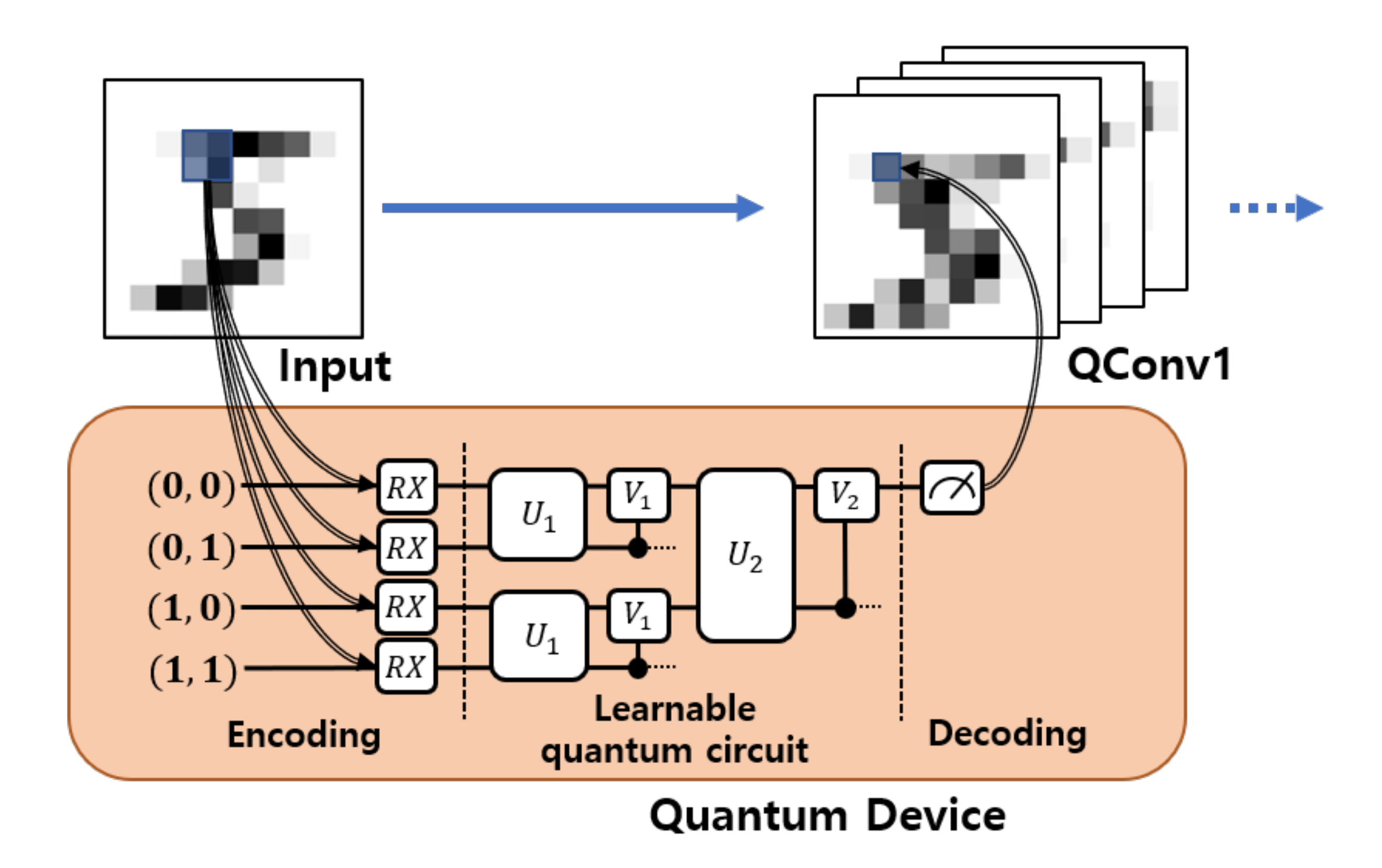}
    \caption{Example of quantum convolution layer for image classification. The learnable quantum circuit can choose in various methods that can enhance performance.}
    \label{fig2}
\end{figure}

The quantum convolution layer can construct, as shown in Fig.~\ref{fig2}.
The concept proceeds as follows:
% quanvolution layer를 구성하기 위해서는 다음과 같은 단계를 따라야 한다
\begin{enumerate}
    \item The encoding process stores the pixel data corresponding to the filter size in qubits.
    % a) N*N의 픽셀 데이터를 인코딩 과정을 통해 N^2개의 qubit의 초기 상태를 결정한다.
    \item The learnable quantum circuits apply the filters that can find the hidden state from the input state.
    % b) input state로부터 학습가능한 random quantum circuit을 이용한 filter를 적용하여 hidden state를 찾아낸다.
    \item The decoding process gets new classical data by measurement.
    % c) 측정을 포함한 decoding 과정으로 새로운 classical data를 얻어낸다.
    \item Repeat steps $1)$ to $3)$ to complete the new feature map.
    % d) a)~d)의 과정들을 반복하여 feature map을 완성시킨다.
\end{enumerate}

The encoding process of $1)$ is a process necessary to convert classical information into quantum information.
% 1)의 인코딩 과정은 고전적인 정보를 양자정보로 변환하기 위해 필요한 과정이다.
The simplest method is to apply a rotation gate corresponding to a pixel data to qubits.
% 가장 간단한 방법으로는 한 픽셀 데이터를 큐빗에 그에 대응하는 크기의 rotation gate를 적용하는 것이다.
Of course, various encoding methods are possible, and the selected encoding method can change the number of qubits required and the learning efficiency.
% 물론 다양한 인코딩 방법이 가능하며, 선택하는 인코딩 방법에 따라 필요한 qubit의 수와, 학습의 효율을 바꿀 수 있다.
The decoding process of $3)$ is determined according to measuring one or more quantum states.
% 3)의 디코딩 과정은 한개 이상의 양자 상태를 측정하는 방법에 따라 결정된다.
By measuring quantum states, classical data are determined. 
% 양자 상태를 측정하여 고전적 정보의 데이터를 결정한다.

The random quantum circuit in $2)$ can be made from a combination of multiple gates.
% 2)의 random quantum circuit은 여러개의 gate의 조합으로 만들어질 수 있다.
Also, the circuit can perform optimization using the gradient descent method by adding variable gates.
% 또한, 이 circuit에 variable gate를 추가하여 gradient descent 방법을 이용해 최적화 학습을 할 수 있다.
This circuit can be designed in various ways that can affect learning performance depending on the design method.
% 이 circuit은 다양한 방법으로 설계될 수 있으며, 설계방법에 따라 학습성능에 차이를 줄 수 있다.
When using MERA, the classical environment generally requires $O(n^2)$ operations in an $n^2$-sized filter, but in a quantum system, the parallelism of qubits can design the filters with $O(log(n))$ depths.
% MERA를 이용할 경우, 고전적 환경에서는 n*n크기의 filter에서 일반적으로 O(n^2)의 연산을 요구하지만, 양자 시스템에서는 큐빗간의 병렬성을 이용하여 O(log(n))의 depth로 filter를 설계할 수 있다.

\section{Learning MNIST Using QCNN Simulation}
% QCNN 시뮬레이션을 이용한 MNIST 학습

In this section, simulations are performed to verify that the actual quantum convolutional neural network works properly in image classification using the MNIST dataset\cite{lecun1998gradient}.
% 이번 섹션에서는 실제 Quanvolution neural network가 이미지 분류에서 제대로 동작하는지 검증하기 위해 시뮬레이션을 진행하였다.
QCNN's quantum computing simulation used the TensorFlow Quantum platform\cite{broughton2020tensorflow}.
% QCNN의 시뮬레이션은 Tensorflow quantum 플랫폼을 이용하였다.
However, because the quantum computing simulation environment uses many resources, it has the following limitations.
% 단, 양자 컴퓨팅 시뮬레이션은 리소스를 많이 사용하기 때문에, 다음과 같이 제한을 두었다.
\begin{itemize}
    \item The $28\cross28$ size MNIST dataset was downscaled to $10\cross10$ size.
    % 기존의 28X28 크기의 MNIST dataset을 10X10 크기로 다운스케일했다.
    \item The filter size of the quantum convolution layer was limited to $2\cross2$.
    % Quanvolution layer의 필터 사이즈를 2X2로 제한하였다.
    \item In each epoch, 2500 random images out of 60,000 are selected for learning.
    % 1회 epoch에서 6만개중의 2500개의 랜덤 이미지를 선택하여 학습을 진행한다.
\end{itemize}
To evaluate the performance of QCNN, fully-connected, CNN, and QCNN models are defined as follows:
% QCNN의 성능을 평가하기 위해, fully-connected, CNN, QCNN 모델을 다음과 같이 정의한다.

\begin{figure}[t]
    \centerline{
        \Qcircuit @C=0.5em @R=0.5em {
        \lstick{(0,0)} & \gate{RX(a_{00})} & \gate{RZ} & \gate{RX} & \gate{RZ} & \gate{RX} & \meter \\
        \lstick{(0,1)} & \gate{RX(a_{01})} & \ctrl{-1} & \ctrl{-1}\\
        \lstick{(1,0)} & \gate{RX(a_{10})} & \gate{RZ} & \gate{RX} & \ctrl{-2} & \ctrl{-2} \\
        \lstick{(1,1)} & \gate{RX(a_{11})} & \ctrl{-1} & \ctrl{-1}\\
        }
    }
    \caption{Circuit example of a simple quantum convolution layer used for learning of the MNIST dataset. The learnable hyperparameters in variable gates optimize the model.}
    % MNIST 학습에 이용된 간단한 quanvolution layer의 circuit 예시.
    % variable gates에 존재하는 학습가능한 hyperparameter를 통해 모델의 최적화가 이루어진다.
    \label{fig3}
\end{figure}
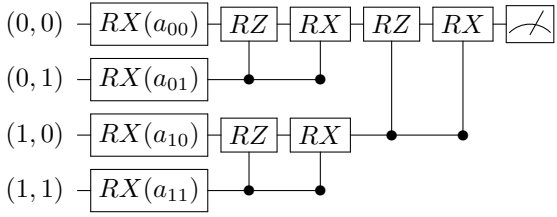

\begin{itemize}
    \item QCNN model: The quantum convolution layer consists of a quantum circuit defined by Fig.\ref{fig3}. The quantum convolution layer returns a feature map with 8 channels. The returned feature map predicts classification results by connecting the feature map with a fully-connected layer comprising 64 and 10 hidden units.
% QCNN model:
% Fig처럼 정의된 quantum circuit으로 이용하는 quantum convolution layer를 이용한다.
% 그 quantum convolution layer는 8개의 채널을 갖는 feature map을 반환한다.
% 반환된 feature map은 feature map을 64개, 10개의 hidden unit로 구성된0 fully-connected layer와 연결하여 분류를 예측한다.
    \item fully-connected model: Construct the model using only the fully-connected layer to check whether the quantum convolution layer affects learning.
% fully-connected model: 
% quantum convolution layer가 실제로 학습에 영향을 주는지 확인하기 위해, 오직 fully-connected layer만을 이용한다.
    \item CNN model: The convolution layer that returns the feature map of the same channel length replaces the quantum convolution layer to compare the performance difference.
% CNN model:
% quantum convolution layer가 convolution layer와의 성능차이를 비교하기 위해, quantum convolution layer와 같은 차원의 feature map을 반환하는 convolution layer로 대체한다.
\end{itemize}

\begin{figure}[t]
    \centering
    \includegraphics[width=\linewidth]{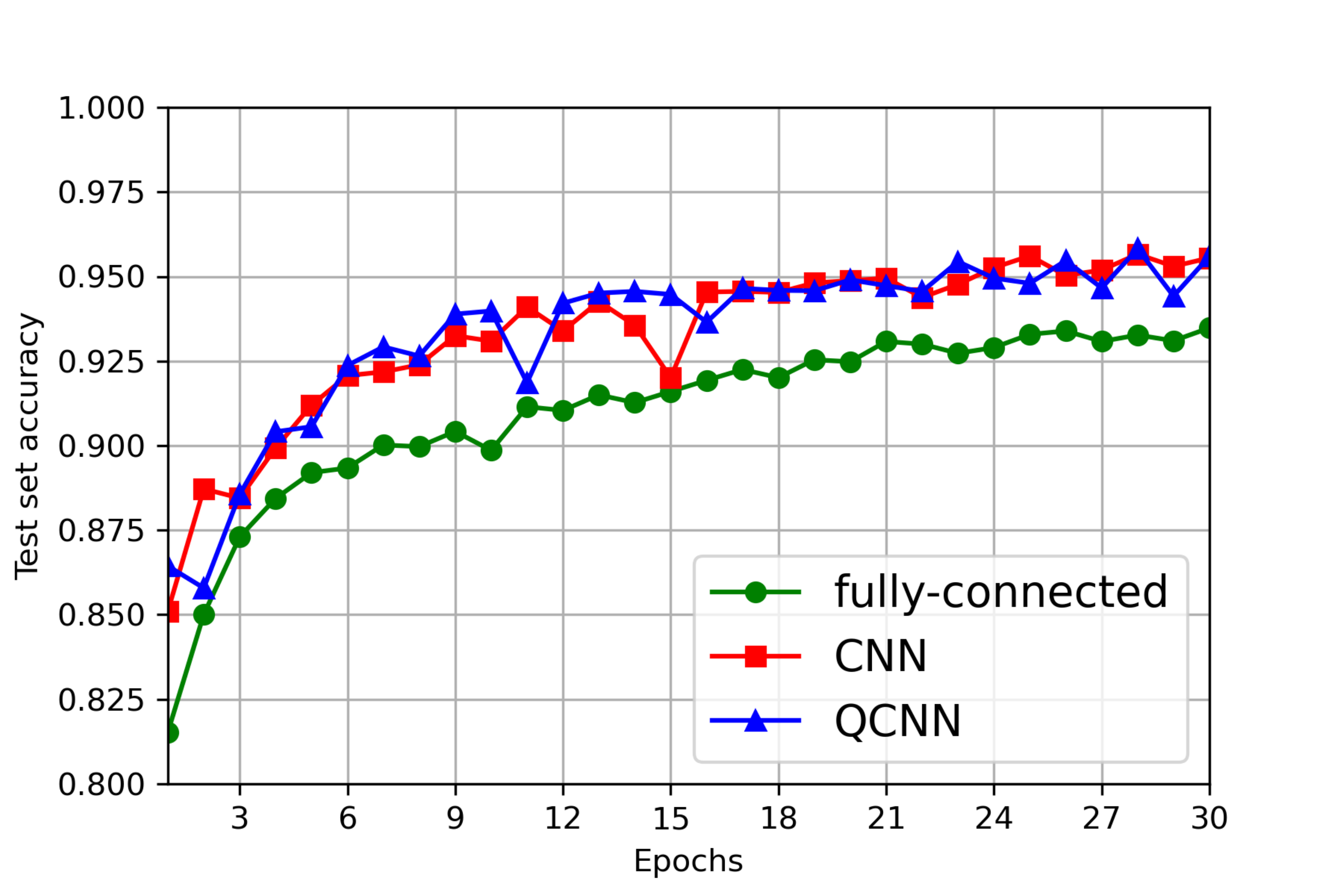}
    \includegraphics[width=\linewidth]{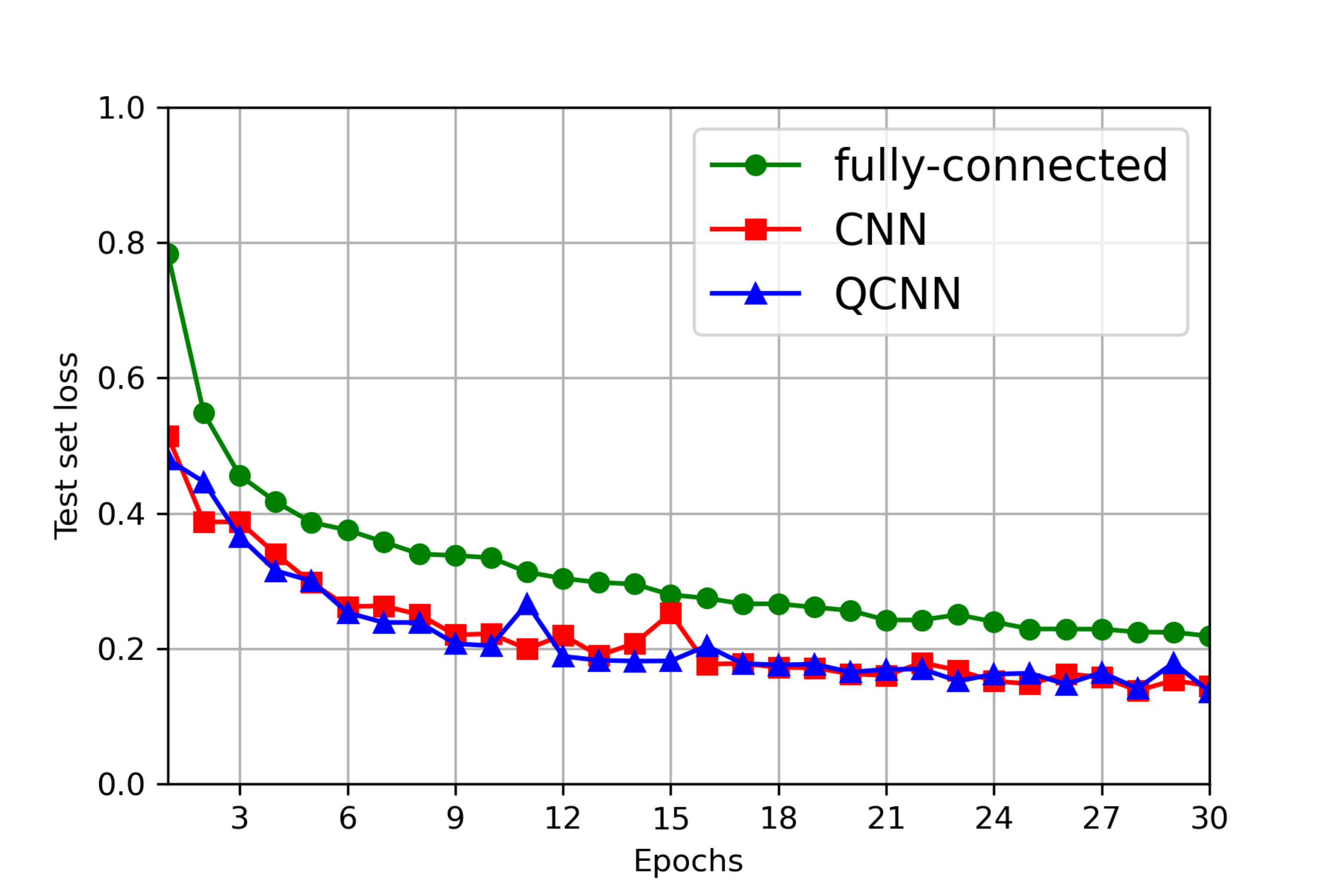}
    \caption{The performance of the QCNN model compared to fully-connected layers only model, and CNN model.}
    % 모델에 따른 학습결과
    \label{fig4}
\end{figure}

Fig.\ref{fig4} shows the results of this simulation.
First, the QCNN model always shows good learning results compared to the fully-connected layer. 
% 첫번째로, Fig.4에서 보이는 그래프와 같이 QCNN model이 fully-connected layer에 비해 모든 결과에서 좋은 학습 결과가 나오는것을 확인할 수 있다.
It confirmed that the quantum convolution layer could improve learning performance. 
% 이는 실제로 quantum convolution layer는 실제로 학습 performance을 올려줄 수 있는것을 확인하였다.
Second, in the comparison between the QCNN model and the CNN model, it can be seen that similar learning results appear. 
% 두번째로, QCNN model과 CNN model과의 비교에서, 서로 비슷한 학습결과가 나오는 것을 확인할 수 있다.
In other words, the QCNN model can have the same learning performance as the CNN model.
% 즉, QCNN model은 QCNN과 동일 학습 퍼포먼스를 가진다.
% 삭제 - NISQ era에서의 quantum computing에서는 보다 효율적인 학습환경을 가질 수 있다.

\section{Concluding Remarks and Future Work}

QCNN uses a CNN model and a quantum computing environment to enable various approaches. 
% QCNN은 CNN 모델과 양자 컴퓨팅 환경을 함께 사용하여, 다양한 접근을 가능하게 해준다.
The QCNN model can be a solution in the field of physical and chemical classification not solved simply, and in addition to the existing CNN model, it can be a more effective and efficient learning model method. 
% 이는 기존에 간단히 해결하지 못한 물리, 화학 분류 분야에서의 해결책이 될 수 있고, 기존 CNN 모델에 추가하여 더 효과적이고 효율적인 학습 모델의 방법이 될 수 있다.
Furthermore, in the quantum computer of the NISQ era, the QCNN model can expect more efficient and high-level results in more complex and large-scale learning\cite{preskill2018quantum}. 
% 그리고 NISQ 시대의 양자컴퓨터에서는 보다 복잡하고 큰 규모의 학습에서도 효율적이고, 높은 수준의 결과를 기대해 볼 수 있을것이다.
In this experiment's results, the simulations were performed at the microscopic scale, but we plan to apply the QCNN model to more complex data.
% 이번 실험 결과에서는 아주 작은 수준에서의 시뮬레이션을 진행했지만, 우리는 보다 복잡한 데이터에서도 QCNN 모델을 적용해볼 계획이다.

QCNN is available in more detailed implementation and approaches. 
% QCNN은 아직 다양한 세부 설계방법과, 접근 방법이 가능하다.
Depending on how the internal quantum circuit is designed, the performance evaluation of the learning model can be improved. 
% 내부 양자 회로를 설계하는 방법에 따라, 학습모델의 성능평가는 더욱 좋아질 수 있다.
Besides, when applying the QCNN model to the field of imaging processing, by including more information in one qubit according to an encoding method, much more efficient learning may be possible. 
% 또한, 이미징 처리 분야에 QCNN 모델을 적용할 때, 인코딩 방법에 따라 큐빗 하나에 더 많은 정보를 포함시키는 것으로, 훨씬 더 효율적인 학습이 가능할 수 있다.
In the future, we will study the QCNN model that is more efficient and has better learning performance through simulation using various approaches.
% 따라서 우리는 미래에 다양한 내부 모델을 이용한 시뮬레이션을 통해, 보다 더 효율적이고, 학습성능이 좋은 QCNN 모델에 대해 연구해볼 것이다.

\section*{Acknowledgment}
This research was supported by National Research Foundation of Korea (2019M3E4A1080391). % 양자과제
J. Kim is a corresponding author (e-mail: joongheon@korea.ac.kr). 

\bibliographystyle{IEEEtran}  
\bibliography{list}

\end{document}